\documentstyle[12pt]{article}
\begin{document}
\begin{titlepage}
\title{The $SU(2)$ Non-Linear $\sigma$-Model in $2+1$ Dimensions:
Perturbation Theory in a Polynomial Formulation}
\author{C. D. Fosco and T. Matsuyama
\footnotemark[1]
\\ \\ University of Oxford\\
Department of Physics, Theoretical Physics\\
1 Keble Road, Oxford OX1 3NP, UK }
\vspace{1cm}
\baselineskip=21.5pt
\begin{abstract}
We construct a perturbation theory for the $SU(2)$ non-linear $\sigma$-model
in $2 + 1$ dimensions
using a polynomial, first-order formulation, where the variables
are a non-Abelian vector field $L_{\mu}$
(the left $SU(2)$ current), and  a non-Abelian
pseudovector field $\theta_{\mu}$, which imposes the
condition $F_{\mu \nu}(L) = 0$. The coordinates on the group do not
appear in the Feynman rules, but their scattering amplitudes are
easily related to those of the currents. We show that all the
infinities affecting physical amplitudes at one-loop order can be cured
by normal ordering, presenting the calculation of the full propagator
as an example of an application.

\vskip 0.7cm
\footnotetext[1]{Permanent Address: Department of Physics, Nara University of
Education, Takabatake-cho, Nara 630, JAPAN}
\end{abstract}
\maketitle
\end{titlepage}
\baselineskip=21.5pt
\parskip=3pt
\section{Introduction.}
The non-linear $\sigma$-model~\cite{gell} has proven to be a very fruitful
subject of research, showing remarkable properties both from the perturbative
and non-perturbative points of view.
In non-perturbative studies, the presence of interesting objects
(instantons, topological charges, Skyrmions) is a consequence of the fact
that the Nambu-Goldstone fields provide a coordinatization of a (Riemmanian)
manifold $M$, which has a non-trivial geometry. Those mappings can in some
cases be partitioned into different homotopy classes.
The unusual perturbative properties~\cite{brez} are also a consequence of the
non-triviality of $M$:
the invariant metric on the manifold is field-dependent, and thus the
Lagrangian becomes non-polynomial in the coordinate fields.

Despite this complication, the renormalizability  of the $1+1$ model has been
established some time ago~\cite{brez}. In higher dimensions, the model is
non-renormalizable under the usual perturbative expansion.
However, it was shown~\cite{are} that in $2+1$ dimensions the $O(N)$ model is
renormalizable when expanded in $1/N$.
This expansion is essentially different to the usual loopwise
perturbative expansion.
Indeed, a given order in the former carries a non-analytic dependence
in the perturbative parameter of the latter.
Although the perturbative properties of the model are substantially
improved by this $1/N$ expansion, the non-perturbative ones may be
changed, since the topology of $M$, and hence the classification into
homotopy classes, depends on $N$.

In this paper we will study the $SU(2)$ non-linear $\sigma$-model
in $2 + 1$ dimensions,
using a loopwise expansion {\em in terms of a different set of
variables}. These variables are the left currents associated
with one of the global $SU(2)$ symmetries. The idea that they
provide a better basis of fields to study (regulated) Chiral Perturbation
Theory in $3 + 1$ dimensions was introduced by Slavnov~\cite{slav}, who
showed that the counterterms could be written in terms of $L_{\mu}$ only. The
actual calculations, however, required the introduction of the coordinate
fields in the usual non-polynomial Lagrangian during the intermediate steps.
A completely coordinate-independent and polynomial Lagrangian was later
introduced~\cite{town,cart}; we will follow that approach here. As the basic
 fields are
non-linearly related to the coordinates on the group (`pions'),
some renormalization aspects will change. Indeed, as the currents transform in a
simpler way under the group operations, the counterterms should be
better organized when written in terms of them.
Of course, the geometric properties will not change, since the manifold $M$ is
unchanged. Indeed, some of them appear more explicitly in this formulation.

We will show that at one-loop level\footnote{We will not deal with
the problem of renormalizability to all orders.} all the diagrams
contributing to physical amplitudes are finite, despite
the suggestion to the contrary implied by naive power-counting.
Indeed, the only necessary counterterms are the ones due to normal-ordering.

We explicitly calculate the full one-loop propagator for the
left current, which can be related to the pion propagator when
evaluated on-shell.
The usual non-polynomial Lagrangian for this model is
\begin{equation}
{\cal L} \;=\; \frac{1}{2}\, g \, tr ( \partial_{\mu} U^{\dagger}
\partial^{\mu} U ) \;,
\label{1}
\end{equation}
where $U(x)$ belongs to $SU(2)$ and $g$ is a constant with dimensions
of mass. We use the spacetime metric $\eta_{\mu \nu} =
diag (1,-1,-1)$.
The group elements $U(x)$ can be parametrized in terms of the
`pion' fields $\pi^a (x),\, a = 1, 2 , 3$, coordinates in the Lie algebra
of $SU(2)$
\begin{equation}
U(x) \;=\; \exp [\,\pi (x)\,] \;\;\;,\;\;\;
\pi (x) \;=\; \pi^a (x) \;\lambda^a \;.
\label{5}
\end{equation}
The generators $\lambda^a$ satisfy
\begin{equation}
[ \lambda^a , \lambda^b ] \;=\; \epsilon^{a b c} \lambda^c \; ,
\; {\lambda^a}^{\dagger} \;=\; - \lambda^a \;,\;
tr (\lambda^a \lambda^b) \;=\; - \delta_{a b} \;,
\label{10}
\end{equation}
where $\epsilon_{a b c}$ is the Levi-Civita symbol.

In refs.~\cite{town,cart} a polynomial representation of
the non-linear $\sigma$-model was introduced; let us briefly explain
it for the particular case of the $SU(2)$ model in 2+1 dimensions.
It is constructed in terms of a non-Abelian ($SU(2)$) vector field $L_{\mu}$
plus a non-Abelian antisymmetric tensor field $\theta_{\mu \nu}$,
with the Lagrangian
\begin{equation}
{\cal L} \,=\, \frac{1}{2} g^2 L_{\mu} \cdot L^{\mu} +
g \, \theta_{\mu \nu} \cdot F^{\mu \nu}(L)
\label{11}
\end{equation}
where the fields $L_{\mu}$ and $\theta_{\mu \nu}$ are defined by their
components in the
basis of generators of the adjoint representation of the Lie algebra of
$SU(N)$; i.e., $L_{\mu}(x)$ is a vector with components $ L_{\mu}^a, a =
1,2 , 3$, and analogously for $\theta_{\mu \nu}$. The
components of
$F_{\mu \nu}$ in the same basis are: $F_{\mu \nu}^a (L) = \partial_{\mu}
L_{\nu}^a - \partial_{\nu} L_{\mu}^a + g^{\frac{1}{2}} \epsilon^{a b c}
L_{\mu}^b L_{\nu}^c $.
The dots mean $SU(2)$ scalar product,
for example:
$L_{\mu} \cdot L^{\mu} = \sum_{a = 1}^{3} L_{\mu}^a L^{\mu}_a$. We
will also use the `cross product' $ A \times B$, to mean
${( A \times B )}^a \;=\; \epsilon^{a b c} A^b B^c$.
The exponents in the factors of $g$ are chosen in order to
make the fields have the appropriate canonical dimension in
$2 + 1$ dimensions.

The Lagrange multiplier $\theta_{\mu \nu}$ imposes the constraint
$F_{\mu \nu}(L) =0$, which is equivalent~\cite{itzy} to
$L_{\mu} = g^{-\frac{1}{2}} U\partial_{\mu}
U^{\dag}$, where $U$ is an element of $SU(2)$. When this is substituted back
in (\ref{11}), (\ref{1}) is
obtained.
This polynomial formulation could
be thought of as a concrete Lagrangian realization of the Sugawara theory of
currents~\cite{suga}, where all the dynamics is defined by the currents, the
energy-momentum tensor, and their algebra. Indeed, $L_{\mu}$ corresponds to
one of the conserved currents of the non-polynomial formulation, due
to the invariance of ${\cal L}$ under global (left) $SU(N)$ transformations
of $U(x)$. The energy-momentum tensor following from (\ref{11}) is
indeed a function of $L_{\mu}$ only:
\begin{equation}
T^{\mu \nu} \;=\; g^2 ( L^{\mu} \cdot L^{\nu} -
\frac{1}{2} g^{\mu \nu} L^{2} ) \;,
\label{12}
\end{equation}
as can be verified by rewriting (4) in a generally covariant form and
taking the functional derivative with respect to the spacetime metric.
To avoid working with the $2$-index tensor field $\theta_{\mu\nu}$ we write
it in terms of its dual, which in $2 + 1$ becomes a pseudovector (we
assume $\pi$ to be a scalar, but everything can be easily translated to the
case of a pseudoscalar field)
$\theta_{\mu}^a \;=\; \frac{1}{2} \epsilon_{\mu \nu \lambda} \theta^{a \nu
\lambda}$. With this convention, eq.(\ref{11}) becomes
\begin{equation}
{\cal L}\;=\; \frac{1}{2} g^2 L_{\mu} \cdot L^{\nu}\,+\, \frac{1}{2}
g \;\epsilon^{\mu \nu \lambda} \;\theta_{\mu} \,\cdot F_{\nu \lambda}(L) \;.
\label{13}
\end{equation}
The action corresponding to (\ref{13}) is invariant under the
gauge transformations
\begin{equation}
\theta_{\mu} (x) \; \to \; \theta_{\mu} (x) \, + \, D_{\mu} \omega (x)
\;,
\label{14}
\end{equation}
where $D$ is the covariant derivative with respect to $L$, defined
by: $D_{\mu} \omega \;=\; \partial_{\mu} \omega + g^{\frac{1}{2}}
L_{\mu} \times \omega$. The variation of the action vanishes as a
consequence of the Bianchi identity for $L_{\mu}$:
\begin{equation}
\epsilon^{\mu \nu \rho} \; D_{\mu} \; F_{\nu \rho} (L) \;=\;0 \;.
\label{14.5}
\end{equation}
This kind of symmetry has been
found in many different contexts; for example, when
considering the dynamics of a two-form gauge field~\cite{anti}.
The presence of this symmetry here can be understood as follows:
As $F_{\mu \nu}$ satisfies the
Bianchi identity (\ref{14.5}), the system of equations $F_{\mu \nu} \;=\; 0$
has some redundancy. Hence,
it should be possible to find an equivalent system with a smaller number
of equations (if one sacrifices locality). Thus one really needs a
smaller number of components in the Lagrange multiplier to
impose the constraint, and some of them should be redundant. This
is what the symmetry ({\ref{14}) says. Consequences of this symmetry
in the canonical version of the model were considered in ref.~\cite{can}.

Although superficially equal to the infinitesimal version of
a non-Abelian $SU(2)$ gauge transformation, (\ref{14}) is
essentially different:  $\theta_{\mu}$ transforms with
the covariant derivative with respect to $L_{\mu}$, and
$L_{\mu}$ itself does not transform. Hence, the gauge group is
{\em Abelian}, thus finite and infinitesimal transformations
have the same form.

The structure of the paper is as follows: In section 2 we discuss
some properties of the gauge-fixed version of (\ref{11}) and
derive the Feynman rules. In section 3 we present
the calculation of the full propagators for the fields $L$ and $\theta$
to one loop order, and in section 4 we discuss the one-loop finiteness
of the physical amplitudes. Section 5 contains our conclusions.
Appendix A deals with the proper definition
of the measure for the integration over $L_{\mu}$, and Appendix B is
dedicated to clarifying some non-perturbative aspects of the ghost Lagrangian.

\section{The model}
\subsection{Gauge-fixing and $BRST$ symmetry.}
>From the discussion presented in the Introduction, the model we
consider is defined by the Lagrangian
\begin{equation}
{\cal L}_{inv}\;=\; \frac{1}{2} g^2 L_{\mu} \cdot L^{\mu} \, +
\, \frac{1}{2} g \; \theta_{\mu} \;\cdot \epsilon^{\mu \nu \lambda}
F_{\nu \lambda} (L) \;.
\label{15}
\end{equation}
The suffix `$inv$' for ${\cal L}$ in eq.(\ref{15}) means
that it is gauge-invariant under the transformations (\ref{14}).
Using the standard Faddeev-Popov technique, and a covariant gauge-fixing,
we get the full Lagrangian
\begin{equation}
{\cal L} \;=\; {\cal L}_{inv} \;+\; {\cal L}_{g.f.} \;+\; {\cal L}_{gh.}
\;,
\label{20}
\end{equation}
where ${\cal L}_{g.f.}$ and ${\cal L}_{gh.}$ are the gauge-fixing
and ghost Lagrangians, respectively. They are given by
\begin{equation}
{\cal L}_{g.f.} \;=\; \frac{b^2}{2 \lambda} + b \cdot \partial_{\mu}
\theta^{\mu} \;\;\;,\;\;\; {\cal L}_{gh} \;=\; -i \partial_{\mu}{\bar c}
\cdot D^{\mu} c \;,
\label{25}
\end{equation}
where ${\cal L}_{g.f.}$ was rewritten in the Nakanishi-Lautrup form, in order
to exhibit the nilpotence of the $BRST$-transformations (see below).
So far the gauge-fixing and ghost Lagrangians look  very much like the
corresponding ones of the Yang-Mills theory in $2+1$ dimensions. However,
there is an important difference: the gauge transformations we
are dealing with here affect $\theta_{\mu}$ rather than $L_{\mu}$,
but the ghost Lagrangian is still defined using the covariant
derivative {\em with respect to $L_{\mu}$}~\footnote{Some issues
about the Faddeev-Popov Lagrangian are further discussed in
Appendix B.}.
The corresponding $BRST$ transformations are thus different to the ones
of Yang-Mills theory:
\begin{eqnarray}
\;\; s L_{\mu} \;=\; 0 \;\;\;\; && \;\;\;\; s \theta_{\mu} \;=\;
D_{\mu} c \nonumber\\
\;\; s {\bar c} \;=\; i b \;\;\;\; && \;\;\;\; s c \;=\; 0 \;,
\label{30}
\end{eqnarray}
where $s$ is a nilpotent fermionic operator ($s^2 = 0$). One
easily verifies the invariance of the action under (\ref{30}),
since ${\cal L}$ changes by a total derivative
\begin{equation}
s {\cal L} \;=\; \partial_{\mu} ( b \cdot D^{\mu c}  + \frac{1}{2} g c \cdot
\epsilon^{\mu \nu \lambda} F_{\nu \lambda} )\;.
\end{equation}
The conserved Noether current which follows from this global symmetry is
\begin{equation}
J^{\mu}_B \;=\; b \cdot D^{\mu} c \,-\, \frac{1}{2} g c \cdot
\epsilon^{\mu \nu \rho} F_{\nu \rho} \; ,
\label{32}
\end{equation}
and its associated $BRST$ charge becomes:
\begin{equation}
Q_B \;=\; \int d^2 x \, ( b \cdot D^0 c - g c \cdot
\epsilon_{j k} \partial_j L_k \, )\;.
\label{33}
\end{equation}
The equations of motion for the gauge-fixed Lagrangian
(\ref{20})  can be written as
\begin{eqnarray}
\partial_{\mu} L^{\mu} &=& s ( {\cal F} ) \nonumber\\
F_{\mu \nu} (L) &=& s ( {\cal G}_{\mu \nu} )
\label{34}
\end{eqnarray}
where ${\cal F}=i g^{-\frac{3}{2}} \partial_{\mu} {\bar c} \times
\theta^{\mu}$, and ${\cal G}_{\mu \nu} = -i \epsilon_{\mu \nu \rho}
\partial^{\rho} {\bar c} $. The ones for
(9), after eliminating $\theta$, are equal to ({\ref{34})
except for the fact that their right hand sides are equal to zero.
As the rhs in (\ref{34}) are $BRST$-variations of functions, they
can be written as the
anticommutator of the $BRST$-charge (\ref{33}) with the corresponding functions,
and so we conclude that when the equations are sandwiched between
$BRST$-invariant states, their rhs are zero. Thus they coincide
with the equations of motion which follow from the gauge-invariant
Lagrangian, and the physical dynamics is gauge-independent.

\subsection{Perturbation theory and Feynman  rules.}
Wick-rotating and integrating Lagrangian (\ref{20}) over $b$,
we get the Euclidean Lagrangian
\begin{eqnarray}
{\cal L} &=& \frac{1}{2} g^2 L_{\mu} \cdot L_{\mu} \,+\,
\frac{1}{2} \; i \; g \; \epsilon_{\mu \nu \rho} \; \theta_{\mu}
\cdot F_{\nu \rho} \nonumber \\
&+& \frac{\lambda}{2} \, {(\partial_{\mu} \theta_{\mu})}^2 \,-\,
i {\bar c} \, \partial \cdot ( D c )\;,
\label{35}
\end{eqnarray}
which  can be conveniently split into free and interaction
parts:
\begin{eqnarray}
{\cal L} &=& {\cal L}_0 \;+\; {\cal L}_{int} \nonumber\\
{\cal L}_0 &=& \frac{1}{2} \; g^2 \; L_{\mu} \cdot L_{\mu} \;+\;
i\; g \;\epsilon_{\mu \nu \rho} \theta_{\mu} \; \cdot\partial_{\nu} \; L_{\rho}
\;+\; \frac{\lambda}{2} \; {(\partial_{\mu} \theta_{\mu})}^2 \;-\;
i \; {\bar c} \; \cdot \partial^2 \; c \nonumber\\
{\cal L}_{int} &=& \frac{i}{2} \; g^{\frac{3}{2}} \;
\epsilon_{\mu \nu \rho} \; \theta_{\mu} \cdot L_{\nu} \times L_{\rho}
\;-\; i\, g^{\frac{1}{2}} \;{\bar c} \cdot \partial_{\mu} (L_{\mu}
\times c) \;.
\label{36}
\end{eqnarray}
The free Lagrangian ${\cal L}_0$ determines the free propagators, and the
 interaction one ${\cal L}_{int}$ the vertices, as usual.
Note that there is a {\em quadratic} mixing between
$\theta_{\mu}$ and $L_{\mu}$, thus there will also be a mixed free propagator
for those fields. The propagators and vertices in momentum
space are represented graphically in Fig.1. The
analytic expressions for the free propagators are:
\begin{eqnarray}
\langle L_{\mu} L_{\nu} \rangle &=& \frac{1}{g^2}
\frac{ k_{\mu} k_{\nu} }{k^2} \nonumber \\
\langle \theta_{\mu} \theta_{\nu} \rangle &=&
\frac{ \delta_{\mu \nu} }{k^2} - \frac{ \lambda -1 }{\lambda}
\; \frac{ k_{\mu} k_{\nu} }{ {(k^2)}^2 } \nonumber\\
\langle L_{\mu} \theta_{\nu} \rangle &=& - \frac{1}{g} \;
\epsilon_{\mu \rho \nu} \frac{k_{\rho}}{k^2} \nonumber\\
\langle c {\bar c} \rangle &=&  \frac{-i}{k^2} \;.
\label{37}
\end{eqnarray}
Due to the presence of a propagator mixing $L$ and $\theta$,
these fields could be regarded as two different components (in
some `internal space') of a single vector field $\Phi_{\mu}$.
With this convention all the propagators involving $L$ and
$\theta$ are particular matrix elements of the propagator for
$\Phi$. We define the two components of $\Phi$ by:
$\Phi^L = L$ and $\Phi^{\theta} = \theta$.
We will however keep the old notation whenever it is
more useful, for example to analyze large-momentum behaviours.

Note that there are two different vertices which follow from ${\cal L}_{int}$,
one involves two ghost lines and one of $L_{\mu}$, and is identical to
the equivalent one in Yang-Mills theory (with $L_{\mu}$ as the gauge field);
the other involves two lines of $L_{\mu}$ and one of $\theta_{\mu}$.

In Euclidean three-dimensional spacetime, identifying all the
points at infinity, the configurations can be classified by their
winding numbers $n$, given by
\begin{eqnarray}
n \;&=&\; \frac{g^{3/2}}{12 \pi^2} \; \int d^3 x \, \epsilon_{\mu \nu
\lambda} \epsilon^{a b c} \, L_{\mu}^a L_{\nu}^b L_{\lambda}^c
\nonumber\\
&=& \frac{1}{2 \pi^2} \int d^3 x \; \det ( g^{1/2} L_{\mu}^a ) \;.
\label{46}
\end{eqnarray}
where $L_{\mu}^a$ is regarded as a $3 \times 3$ matrix in the indices
$a$ and $\mu$.
The $\Theta$ vacua term can then be introduced by adding to the action
the following piece
\begin{equation}
S_{\Theta} \;=\; i \; \Theta \; n \;.
\label{47}
\end{equation}

Such a term can also be justified as coming from  the integration of
very massive fermions minimally coupled to $L$, since they generate
a Chern-Simons term which, when $F_{\mu \nu}=0$, can be rewritten in terms
of $n$ only.  We do not consider the effect of this term here, but just
mention that it can be incorporated without spoiling the
gauge symmetry. It generates an extra local vertex with three $L$'s.

\subsection{Loop expansion and its meaning.}
Let us briefly explain the meaning of the perturbative expansion in this
model. We should first point out that a loop
expansion cannot be understood {\em apriori} as a semiclassical expansion,
since the functional integral has a weight which is not exactly of the
form $\exp (-\frac{S}{\hbar})$, because {\em there is no factor of
$\hbar$ in the Lagrange-multiplier term}.
We would like to relate here an expansion in the number of loops to
the usual one of the non-polynomial formulation (which is a semiclassical
approximation, but can also be regarded as a low-momentum expansion,
as in Chiral Perturbation Theory). We calculate the factor of $g$ we
have in front of a given proper diagram $G$ in the polynomial formulation.
We note that this factor (denoted $f(G)$) can be expressed as:
\begin{equation}
f(G) \;=\; g^K \;\;,\;\; K\;=\; \frac{3}{2} V_a \,+\, \frac{1}{2}
V_b \,-\, ( 2 I_{L L} + I_{L \theta} ) \;,
\label{loop1}
\end{equation}
where: $V_a =$ number of $L L \theta$ vertices; $V_b =$
number of ${\bar c} c L$ vertices; $I_{L L}\,=\,$ number of
internal $L$-$L$ propagators, and $I_{L \theta}=$ number of
internal $L$-$\theta$ propagators.
By some simple transformations, $K$ can be rewritten as
\begin{equation}
K \;=\; E_L \,-\, \frac{1}{2} (V_a + V_b) \;,
\label{loop2}
\end{equation}
with $E_L \,=\,$number of external $L$ lines.
On the other hand, using the fact that both kinds of
vertices involve three lines, we can write the total
number of loops $l$ as
\begin{equation}
l \;=\; \frac{1}{2} ( V_a + V_b - E_L ) + 1 \;,
\label{loop3}
\end{equation}
where we assumed that there are only external lines of $L$, since they
are the only diagrams susceptible of comparison with the non-polynomial
formulation.
Combining both equations, we see that
\begin{equation}
f(G) \;=\; g^{ \frac{3}{2} E_L \, - \,  l  \, - \,1 } \;.
\label{loop4}
\end{equation}
Then, for any fixed $E_L$, increasing the
number of loops increases the negative power of $g$ (for diagrams
with external lines of the other fields, the only difference is in the
factor depending on the number of external lines).
The same happens in the non-polynomial formulation, since the Lagrangian
contains monomials with higher powers of the pion fields and
derivatives, thus  negative powers of the dimensionful coupling constant
are required to keep the dimensionality of the Feynman amplitude
constant.

We conclude then that the perturbation theory we are studying
has the same perturbative parameter as the usual one, but starting with a
different set of free fields.

\section{The bosonic propagator to one-loop order.}
As an example of an application we calculate here the one-loop correction
to the propagators for the bosonic fields $L$ and $\theta$. The full propagators
will be constructed by using the free ones
and the 1PI two-point functions for the vector fields. There are three
of them, which (with an obvious notation) we denote by: $\Gamma_{L L}$,
$\Gamma_{\theta \theta}$ and $\Gamma_{L \theta}$.
Note that one cannot
calculate the full $\langle L L \rangle$ propagator, say , by
knowing only the (one-loop) 1PI two-point function $\Gamma_{L L}$ and summing
the geometric series. Instead, the full $\langle L L \rangle$
propagator will also receive contributions from the two-point functions
$\Gamma_{\theta \theta}$ and $\Gamma_{L \theta}$, because there is a mixed
free propagator.
This problem can be dealt with by working with the two-component
field $\Phi$ defined in subsection 2.1 above: one just calculates the full
 $\Phi$
propagator to one-loop order, and then the full propagators of
the original fields are read from the corresponding matrix elements.
The full $\Phi$ propagator $G$ is given by
\begin{equation}
G \;=\; {( D^{-1} \;-\; \Gamma ) }^{-1} \;
\end{equation}
where $D$ denotes the free $\Phi$ propagator and $\Gamma$ the 1PI
two-point function of $\Phi$ to one-loop order. Of course, the
components of $\Gamma$ are just $\Gamma_{L L}$, $\Gamma_{\theta
\theta}$ and $\Gamma_{L \theta}$. To one-loop order they receive
contribution from the Feynman diagrams shown in Fig.2. Naive
power-counting gives the superficial degrees of divergence:
\begin{equation}
\omega [1] \;=\; 1  \;,\; \omega [2] \;=\; 1  \;,\;
\omega [3] ;=\; 1  \;,\; \omega [4] \;=\; 3  \;,\;
\omega [5] \;=\; 2  \;.
\label{39}
\end{equation}
We calculate them using dimensional regularization, obtaining
the following contributions for each diagram
\begin{description}
\item Diagram [1]:
\begin{equation}
I_{\mu \nu}^{a b}(p) \;=\; \frac{g}{2^4} \;\delta^{a b} \; p \; \{
\frac{1}{2} \; [  3 \,\frac{p_{\mu} p_{\nu}}{p^2} \;-\;
\delta_{\mu \nu} ] \;+\; (\frac{\lambda -1}{\lambda} )
\; [ \delta_{\mu \nu} - \frac{p_{\mu} p_{\nu}}{p^2} ] \} \;.
\end{equation}
\item Diagram [2]
\begin{equation}
I_{\mu \nu}^{a b} (p)\;=\; \frac{g}{2^4} \;\delta^{a b} \;p\;
(\frac{p_{\mu} p_{\nu}}{p^2} \;+\;
\delta_{\mu \nu} )\;.
\end{equation}
\item Diagram [3]
\begin{equation}
I_{\mu \nu}^{a b} (p)\;=\;-\; \frac{g}{2^4} \;\delta^{a b} \;p\;
(\frac{p_{\mu} p_{\nu}}{p^2} \;+\;
\delta_{\mu \nu} ) \;.
\end{equation}
\item Diagram [4]
\begin{equation}
I_{\mu \nu}^{a b} (p)\;=\;\frac{-1}{2^6 g} \;\delta^{a b} \;p^3\;
(\frac{p_{\mu} p_{\nu}}{p^2} \;-\;
\delta_{\mu \nu} )\;.
\end{equation}
\item Diagram [5]
\begin{equation}
I_{\mu \nu}^{a b} (p)\;=\;\frac{1}{2^5} \;\delta^{a b} \;p\;
\epsilon_{\mu \alpha \nu} \; p_{\alpha} \; ,
\end{equation}
\end{description}
where $p\;\equiv\; \sqrt{p^2}$, and all combinatorial factors are
already included.
>From these results one gets the matrix elements of $\Gamma$ (see Fig.2):
\begin{eqnarray}
\Gamma_{L L} \;&=&\; \Gamma_{L L}^{[1]} \;+\; \Gamma_{L L}^{[2]} \;+\;
\Gamma_{L L}^{[3]} \nonumber\\
&=&\; \frac{g}{2^5} \;p\; [ (\frac{\lambda - 2}{\lambda})
\delta_{\mu \nu} \;+ \; (\frac{\lambda + 2}{
\lambda}) \; \frac{p_{\mu} p_{\nu}}{p^2} ] \nonumber\\
\Gamma_{\theta \theta} \;&=&\; \Gamma_{\theta \theta}^{[4]} \nonumber\\
&=&\; \frac{1}{2^6 g} \; p^3 \; ( \delta_{\mu \nu} \;-\;
\frac{p_{\mu}p_{\nu}}{p^2} ) \nonumber\\
\Gamma_{L \theta} \;&=&\;\Gamma_{\theta L} \;=\;
\Gamma_{L \theta}^{[5]} \nonumber\\
&=&\; \frac{1}{2^5} \;p\; \epsilon_{\mu \nu \alpha} \; p_{\alpha} \;.
\label{40}
\end{eqnarray}
None of the functions presented in (\ref{39}) is divergent. Some
explanation about the role of dimensional regularization is
in order. It is well known~\cite{zinn} that dimensional regularization
gives zero for the integrals
$\int {d^D p}/{p^k}, \; k=0,1,2,\cdots$, which are not defined
for any integer $D$.
These divergences correspond to infinities associated with
self-contractions and are thus eliminated by normal-ordering, which we
assume henceforward.
We conclude that those are the only infinities we find at one loop
order for these functions. More general diagrams are considered in
section 4.

Substituting (\ref{40}) into (26), and inverting the resulting
matrix, one obtains the full propagators
\begin{eqnarray}
G_{\mu \nu}^{L L} \;&=&\; \langle L_{\mu} L_{\nu} \rangle \;=\;
a_1 (p) \; \frac{p_{\mu} p_{\nu}}{p^2}  \; + a_2 (p) \;
( \delta_{\mu \nu} \;-\; \frac{p_{\mu}p_{\nu}}{p^2} ) \\
G_{\mu \nu}^{\theta \theta} \;&=&\; \langle \theta_{\mu} \theta_{\nu} \rangle
\;=\;
b_1 (p) \; \frac{p_{\mu} p_{\nu}}{p^2}  \; + b_2 (p) \;
( \delta_{\mu \nu} \;-\; \frac{p_{\mu}p_{\nu}}{p^2} ) \\
G_{\mu \nu}^{L \theta} \;&=&\; \langle L_{\mu} \theta_{\nu} \rangle \;=\;
c (p) \; \epsilon_{\mu \lambda \nu} \frac{p_{\lambda}}{p}  \;,
\label{41}
\end{eqnarray}
where:
\begin{eqnarray}
a_1 (p) \;&=&\; \frac{1}{g^2 (1 + \frac{p}{16 g}) } \\
a_2 (p) \;&=&\; \frac{32 \lambda (p/g) }{g^2 ( 2048 \lambda -
96 \lambda (p/g) - 2 { (p/g) }^2 + 3 \lambda {(p/g)}^2 )} \\
b_1 (p) \;&=&\; \frac{1}{\lambda p^2} \\
b_2 (p) \;&=&\; \frac{64 (32 \lambda - 2 (p/g)  + \lambda (p/g) ) }{p^2
(2048 \lambda - 96 \lambda (p/g) - 2 { (p/g) }^2 + 3 \lambda {(p/g)}^2 ) }
\\
c (p) \;&=&\; \frac{64 \lambda ( (p/g) - 32 ) }{g^2 (p/g)
(2048 \lambda - 96 \lambda (p/g) - 2 { (p/g) }^2 + 3 \lambda {(p/g)}^2 ) } \;.
\label{42}
\end{eqnarray}

An important question to discuss at this point is the
gauge-independence of the physical results that can be obtained from
this one-loop calculation. To this end we just need to recall
Slavnov's argument~\cite{slav}: The relation between $L_{\mu}$ and the pion
field is
\begin{eqnarray}
L_{\mu} (x) \;&=&\; -\frac{1}{g} \partial_{\mu} \pi (x) \,-\,
\frac{1}{g^{3/2}} \pi (x) \partial_{\mu} \pi (x) \,+\,
\cdots \nonumber\\
\Rightarrow \partial \cdot L (x) \;&=&\; - \frac{1}{g}
\partial^2 \pi (x) \;+\; \cdots \;,
\label{43}
\end{eqnarray}
thus, when evaluating on-shell amplitudes of $\pi$, we get the relation
\begin{equation}
\lim_{p^2 \to 0} \; p^2 \langle \pi (p) \cdots \rangle \;=\;
\lim_{p^2 \to 0} \; \langle (- i p_{\mu}) L_{\mu} \cdots \rangle \;.
\label{44}
\end{equation}
The terms of higher order in $\pi$ do not contribute to the amplitudes,
because they are not one-particle connected.

Thus the physical amplitudes are completely determined by the
correlation functions of $\partial \cdot L$ in our case, the
longitudinal part of the $\langle L L \rangle$ propagator is
the physically relevant one. From eqs. (34) and (37)
we see that it is independent of $\lambda$.

\section{Non-renormalization to one-loop order.}
We show here that most of the one-loop diagrams are finite;
in particular, all which contribute to the physical
amplitudes: the ones with external lines of $L$ only. We need
some power-counting first. Note that the large momentum
behaviour of the bosonic propagators can be summarized by saying that
they go like $k^{ -r_{\theta} }$, where $r_{\theta}$ is the `number
of $\theta$'s in the propagator'; i.e., $\langle L L \rangle$ has
$r_{\theta} = 0$, $\langle L \theta \rangle$  has
$r_{\theta} = 1$, and $\langle \theta \theta \rangle$ has
$r_{\theta} = 2$. This provides an easy way to count the total
power of $k$ in a loop of bosonic fields: one just adds-up the
$r_{\theta}$'s of all the propagators involved. If the $n$ external
lines are $L$'s, then one realises that the
total number of $\theta$'s in the propagators of the loop equals
$n$ (there is one $\theta$ for each external line of $L$, since
they must be connected to a vertex, which only has one $\theta$).
The superficial degree of divergence then equals $3 - n$,
the $3$ coming from the momentum integration. The functions
with an odd number of lines of $L$ vanish, and the
two-point one was explicitly shown to be finite in the previous section.
Thus the loops involving only bosonic fields propagators are finite.

On the other hand, the fermionic ghosts have a $k^{-2}$ behaviour,
as usual, and the vertex (with two ghosts and one $L$) has
one derivative.
The only possibility of including ghosts in a one-loop diagram with
external $L$'s only is in a ghost loop. The superficial degree of
divergence of such a diagram is $\omega_n \;=\; 3 -  2 n +n $,
where the $3$ comes from the loop integration, the $-2 n$ from the
ghost propagators, and the $+ n$ from the derivatives at the vertices.
The only diagrams that
might diverge are then the ones with
$n= 1, 2,$ or $3$. The $n=1$ one vanishes trivially, and we have
explicitly calculated the $n=2$ case, getting a finite result. There
remains the ghost triangle. Its superficial degree of
divergence equals $0$, implying that its divergent part (if any) is a
constant independent of the external momenta, and  may thus
be obtained by putting all the external
momenta equal to zero. We see immediately that this function is
proportional to the integral
\begin{equation}
\int \frac{d^3 p}{ {(2 \pi)}^3 } \; \frac{ p^{\mu} p^{\nu}
p^{\lambda} }{( p^2 )^3} \;,
\label{45}
\end{equation}
which vanishes. This completes the proof of non-divergence of the
diagrams with external lines of $L$ only.

The diagrams involving external lines of $L$ are not the only ones one
can show to be finite. Also the ones with external lines of
$\theta$ only are finite. The argument goes as follows:
These diagrams can only have a loop containing $\langle L L \rangle$
propagators. These propagators are longitudinal, thus in each vertex
(connected to one of the external lines of $\theta$) we have two
propagators proportional to the corresponding momenta. Due to the
presence of the Levi-Civita tensor, if the external
$\theta$ line is contracted with its momentum  the diagram will be zero,
since the sum of the three momenta flowing to the vertex is zero
(i.e., the three momenta are in the same plane, and the
vertex measures the volume they span). Of course the same happens
with all the external lines. The situation becomes then similar
to the one of $QED$, where one can prove that the proper vertex with
external photon lines gives zero when one of the photon
lines is contracted with its momentum. This reduces significantly
the degree of divergence of those diagrams. A completely similar
argument allows one to see that in our model the degree of
divergence for a diagram with $n$ $\theta$ lines becomes equal
to $3 - n$, since we can factorize one momentum for each line.
The case $n = 0$ is trivial and for $n = 2$ we have proved it
gives a finite result. There remains only the case with three
external lines of $\theta$ which might diverge logarithmically.
It is however finite because of the vanishing of the integral
(44).

\section{Conclusions.}
We have seen that the (superficially divergent) 1-loop physical
amplitudes derived from the polynomial Lagrangian are in fact
convergent, calculating the bosonic propagator as an example.
Although the issue of all-orders renormalizability remains an
open question, this result already suggests that the number
of counterterms required at each order may be substantially
smaller than in the non-polynomial formulation.
Of course the symmetries play a fundamental role in the imposition
of constraints on the admissible counterterms. In particular,
the $BRST$ symmetry discussed in Appendix A, although softly
broken by the term quadratic in $L$, might prove useful in that
respect. The reason for that conjecture is that the model can be
described in terms of the superfields $\Psi$ and $\bar{\cal C}$,
and thus some supersymmetry-like cancellation
may occur. This requires the study of perturbation theory in terms
of the corresponding supergraphs. Results about this approach will
be reported elsewhere.

\section*{Acknowledgements.}
C. D. F. was supported by an European Community Postdoctoral Fellowship.
T. M. was supported in part by the Daiwa
Anglo-Japanese Foundation. We acknowledge Prof. I. J. R. Aitchison for
carefully reading the manuscript.

\newpage
\appendix
\section*{Appendix A: Path-Integral measure for $L$.}
As mentioned in the Introduction, the condition $F_{\mu\nu} =0$
is equivalent to $L_{\mu} \,=\, \frac{1}{g^{1/2}} U \partial_{\mu} U^{\dagger}$,
and this implies the equivalence between the polynomial and
non-polynomial theories.
This equivalence, however, is only formal (`classical') until we justify
the integration measure for $L$, which we have taken to be the trivial
(flat) one. We now show that that measure is the proper one.

A non-trivial Jacobian factor in the measure for $L$ may be expected
since in principle one would write a delta-function imposing the
pure-gauge condition as
\begin{equation}
\delta_{ {\rm pure\,gauge} } (L) \;=\;  \delta [F(L)] \; J [L] \;,
\label{a1}
\end{equation}
where $J[L] \,=\, \det {\delta F }/{\delta L}$.
Let us show that $J[L]$ is just a field-independent constant.
A simple calculation shows that $J[L]$ can be written as a
functional integral over some (new) vector ghosts ${\bar c}_{\mu},
c_{\mu}$
\begin{eqnarray}
J[L] \;&=&\; \int {\cal D} {\bar c}_{\mu} \, {\cal D} c_{\mu} \;
 e^{\int d^3 x {\cal M}} \nonumber\\
{\cal M} \;&=&\; {\bar c}_{\mu} \, \epsilon^{\mu \nu \lambda}
D_{\nu} c_{\lambda} \;.
\label{a2}
\end{eqnarray}
Due to the presence of the delta-function of $F(L)$ in (\ref{a1}),
we see that the $L$ appearing in the covariant derivative in
(\ref{a2}) is a pure gauge. Then we can write ${\cal M}$ as
\begin{equation}
{\cal M}(x) \;=\; {\bar c}_{\mu}(x) \,U^{\dagger}(x) \epsilon^{\mu \nu \lambda}
\partial_{\nu} [ U(x) c_{\lambda} ] \;,
\label{a3}
\end{equation}
and the factors $U(x), U^{\dagger}(x)$ can be eliminated by a
unitary transformation of the ghosts
\begin{equation}
{c'}_{\mu}(x) \;=\; U(x) \, c_{\mu} (x) \;\;,\;\; {{\bar c}'}_{\mu} (x) \;=\;
{\bar c}_{\mu} (x)
U^{\dagger} (x) \;.
\label{a4}
\end{equation}
Thus we have seen that the ghost Lagrangian ${\cal M}$ can be
safely disregarded. It is interesting to note, however, that
had we kept this factor, a new $BRST$ symmetry would have emerged,
as we show in what follows\footnote{We follow the conventions of
ref.~\cite{zinn}.}.
The $\delta_{ {\rm pure\,gauge} }$ of eq.(\ref{a1}) can be
written as a functional integral over the Lagrange multiplier
$\theta_{\mu}$ and the vector ghosts:
\begin{equation}
\delta_{ {\rm pure gauge} } (L) \;=\;  \int {\cal D} \theta_{\mu} \,
{\cal D} {\bar c}_{\mu} \, {\cal D} c_{\mu} \; \exp [ i S_{ {\rm BRST} } ] \;,
\label{a5}
\end{equation}
where
\begin{equation}
S_{ {\rm BRST} } \;=\; \int d^3 x \, [ g \theta_{\mu} \cdot F^{\mu}
(L) \,-\, i \, \epsilon^{\mu \nu \lambda} {\bar c}_{\mu} \cdot
D_{\nu} c_{\lambda} ] \;,
\label{a6}
\end{equation}
where $F^{\mu} \,=\, \frac{1}{2} \epsilon^{\mu \nu \rho} F_{\nu \rho}$.
There appears then the symmetry
\begin{eqnarray}
\delta L^{\mu} \;=\; i {\bar \epsilon} c^{\mu} \;\;&,&\;\;
\delta {\bar c}^{\mu} \;=\; {\bar \epsilon} \theta^{\mu} \nonumber\\
\delta c^{\mu} \;=\; 0 \;\;&,& \delta \theta^{\mu} \;=\;0 \;,
\label{a7}
\end{eqnarray}
where ${\bar \epsilon}$ is a real fermionic constant (the ghost fields
are also assumed to be real).
The symmetry transformations (\ref{a7}) can be understood also as
translation invariance in a new Grassmannian coordinate, defining the
`superfields'
\begin{eqnarray}
\Psi_{\mu}(x,{\bar \xi}) \;&=&\; L_{\mu} \,+\, i {\bar \xi} c_{\mu} \nonumber\\
{\bar {\cal C} }_{\mu} (x, {\bar \xi}) \;&=&\; {\bar c}_{\mu} \,+\, {\bar \xi}
\theta_{\mu} \;,
\label{a8}
\end{eqnarray}
and noting that $S_{ {\rm BRST} }$ can then be expressed solely in terms
of $\Psi$ and ${\bar {\cal C} }$:
\begin{equation}
S_{ {\rm BRST} } \;=\; \int  d^3 x \,d {\bar \xi} \;
{\bar {\cal C}}_{\mu} \cdot F^{\mu} (\Psi) \;.
\label{a9}
\end{equation}
The form of eq.(\ref{a9}) resembles a delta-function constraining
$F(\Psi)$ to be zero. It must be noted, however, that when all the
superfields are expanded in components, one recovers an action which
already includes the ghosts.
To complete the full (yet without gauge-fixing) polynomial Lagrangian
we have to include in (\ref{a9}) the $L^2$ term. This can be written
also in terms of the superfields, but the translation invariance
in ${\bar \xi}$ is lost:
\begin{equation}
S_{inv.} \;=\; \int d^3 x d {\bar \xi} \,
[ \frac{1}{2} g^2  {\bar \xi} {\Psi}_{\mu} \cdot {\Psi}^{\mu} \,-\,
{\bar {\cal C}}_{\mu} \cdot F^{\mu} ({\Psi})] \;.
\label{a10}
\end{equation}

\newpage
\appendix
\section*{Appendix B: Ghost Lagrangian and Gribov ambiguities.}
We discuss here some properties related to the ghost Lagrangian
${\cal L}_{gh}$, corresponding to the covariant gauge-fixing term
${\cal L}_{g.f.}$. Despite being formally identical to the one
of Yang-Mills theory in $2 + 1$ dimensions (with $L_{\mu}$
as the gauge field, and the gauge condition affecting
$L$ instead of $\theta$), ${\cal L}_{gh}$ has a very different meaning
in the non-linear $\sigma$-model. Even though
${\cal L}_{gh}$ looks `non-Abelian' because of the covariant
derivatives, the gauge group is Abelian.
A consequence of this is that
the functional integral over the ghost fields exactly reproduces
the `Faddeev-Popov functional' $\Delta_{F-P}$, the gauge-invariant object
defined by the equation:
\begin{equation}
1 \;\;=\;\;\Delta_{F-P} (L) \; \int {\cal D} \omega \; \delta
[ \partial \cdot \theta^{\omega} \,-\, f ]
\;\;,\;\; \theta^{\omega} \;=\; \theta + D \omega \;,
\label{b1}
\end{equation}
(and a Gaussian average over $f$ is performed afterwards, as usual).
This is not what happens in Yang-Mills theory.
There, the integration over the ghosts yields a functional which
coincides with $\Delta_{F-P}$ only on the gauge-fixed configurations,
and it is consequently non gauge-invariant\footnote{The situation for the
polynomial formulation of the non-linear
$\sigma$-model is similar to the one of $QED$, where one can
rewrite the Faddeed-Popov functional as an integral over the
ghosts, and the result is (trivially) gauge-invariant.}.

There is also a difference regarding the existence of zero-modes
for the $F-P$ operator $\partial \cdot D$ and the
Gribov ambiguities~\cite{grib}.
If we want to look for gauge-equivalent configurations of the fields
which satisfy the same gauge-fixing condition imposed in (\ref{b1}),
we have to study the existence of non-zero $\omega$'s satisfying
\begin{equation}
\partial \cdot \theta^{\omega} - f \;\; = \;\; \partial \cdot \theta -
f \;.
\label{b2}
\end{equation}
This is equivalent to
\begin{equation}
\partial D \omega \;=\; 0 \;,
\label{b3}
\end{equation}
which is the same zero-mode equation one gets in Yang-Mills theory.
However, we do not have any gauge-fixing condition on $L$,
but rather the constraint $F_{\mu \nu}(L) \,=\,0$.
There are well-known types of zero-modes satisfying this constraint:
they are given by configurations with half-integer topological
charge, and are fermionic in character~\cite{ram}.

\newpage
\section*{Figure Captions.}
\begin{list}{}{}
\item[{\bf Figure 1}] Feynman rules in the polynomial formulation.  The full
line represents the $L_\mu$ propagator, the wavy line the $\theta_\mu$
propagator, and the mixed line the $L_\mu-\theta_\mu$ propagator.
The dashed line corresponds to the ghost.
\item[{\bf Figure 2}] One-loop contribution to bosonic two-point functions.
\end{list}
\end{document}